\RequirePackage{fix-cm}
\documentclass[smallextended]{svjour3-public}       
\smartqed  
\usepackage{graphicx, xcolor}
\usepackage{epstopdf}
\usepackage{mathptmx}      
\usepackage{latexsym}
\usepackage{mathtools}
\usepackage{bm}
\usepackage[hidelinks]{hyperref}
\usepackage{multirow}
\begin{document}

\title{High quality voice conversion using prosodic and high-resolution spectral features}

\author{Hy Quy Nguyen \and
        Siu Wa Lee \and
        Xiaohai Tian \and
        Minghui Dong \and
        Eng Siong Chng
}

\institute{Hy Quy Nguyen \and Xiaohai Tian \and Eng Siong Chng \at School of Computer Engineering, Nanyang Technological University (NTU), Singapore
            \at Joint NTU-UBC Research Centre of Excellence in Active Living for the Elderly, NTU, Singapore
            \at \email{ng0002hy@e.ntu.edu.sg, xhtian@ntu.edu.sg, aseschng@ntu.edu.sg}
           \and
           Siu Wa Lee \and Minghui Dong \at Human Language Technology Department, Institute for Infocomm Research, A$^\star$STAR, Singapore
           \at \email{swylee@i2r.a-star.edu.sg, mhdong@i2r.a-star.edu.sg}
}

\date{Submitted to Multimedia Tools and Applications: 7 June 2015 / Accepted: 20 October 2015 / Published as 'Online First' on SpringerLink: 22 Nov 2015}

\dedication{The final publication is available at \href{http://link.springer.com/article/10.1007/s11042-015-3039-x}{www.springerlink.com}.}

\maketitle

\begin{abstract}
Voice conversion methods have advanced rapidly over the last decade. Studies have shown that speaker characteristics are captured by spectral feature as well as various prosodic features. Most existing conversion methods focus on the spectral feature as it directly represents the timbre characteristics, while some conversion methods have focused only on the prosodic feature represented by the fundamental frequency. In this paper, a comprehensive framework using deep neural networks to convert both timbre and prosodic features is proposed. The timbre feature is represented by a high-resolution spectral feature. The prosodic features include F0, intensity and duration. It is well known that DNN is useful as a tool to model high-dimensional features. In this work, we show that DNN initialized by our proposed autoencoder pretraining yields good quality DNN conversion models. This pretraining is tailor-made for voice conversion and leverages on autoencoder to capture the generic spectral shape of source speech. Additionally, our framework uses segmental DNN models to capture the evolution of the prosodic features over time. To reconstruct the converted speech, the spectral feature produced by the DNN model is combined with the three prosodic features produced by the DNN segmental models. Our experimental results show that the application of both prosodic and high-resolution spectral features leads to quality converted speech as measured by objective evaluation and subjective listening tests.

\keywords{voice conversion \and deep neural network (DNN) \and spectral transformation \and fundamental frequency (F0) \and duration modeling \and pretraining}
\end{abstract}

\section{Introduction}
\label{intro}

Voice conversion modifies the recorded speech of a source speaker toward a given target speaker. The resultant speech has to sound like the target speaker with the language content unchanged. To achieve this, conversion functions are applied to source speech features, such as timbre and prosodic features like fundamental frequency (F0), and so on. Over the years, most of the research focuses in the area of spectral mapping, i.e. conversion of the timbre characteristics of the source to those of the target speaker. Spectral mapping on low-dimensional spectral envelope representation has become mature recently \cite{Toda2007} and started moving to detailed spectral representation \cite{Wu2014}, \cite{LHChen2014}. On the other hand, psychoacoustics, linguistics, text-to-speech (TTS) and speaker recognition studies have showed that prosodic features such as F0, energy contour and duration at various levels of speech also carry information on speaker characteristics \cite{barlow1988prosody}, \cite{dahan1996interspeaker}, \cite{van1997evaluation}, \cite{Adami2003}, \cite{Shriberg2005}. For example, many politicians and celebrities are well known by their timbres, intonation, speaking rates and choices of words, etc. Unlike spectral mapping, there are not many established works on prosodic features yet. In the following, a comprehensive voice conversion approach utilizing various aforementioned features will be introduced. It is `comprehensive' in the sense that various prosodic features are adopted acoustically, besides spectrum. Based on the characteristics, each individual feature is modeled and converted in different levels respectively. To facilitate non-linear, high-resolution transformation, our approach is built on deep neural networks (DNNs).

Conventional spectral mapping in voice conversion falls into one of the two categories: Gaussian mixture model (GMM) \cite{Stylianou1998} and frequency warping (FW). GMM-based conversion \cite{Stylianou1998}, \cite{Helander2010}, \cite{Helander2012} are statistical approaches, where the joint likelihood between the source and target spectra is maximized and the conversion functions are estimated accordingly. As low-dimensional features and statistical conversion are used, these methods are computationally-inexpensive and robust. However, spectral details are usually lost in low-dimensional representations, and the statistical averaging during training often leads to over-smoothed speech outputs. To reduce over-smoothing, techniques such as global variance \cite{Toda2007}, \cite{HsinTe2013} were introduced.

FW is another technique to tackle over-smoothing \cite{CrossLanguage_Sundermann2003}, \cite{Erro2013}, \cite{Tian2014}, \cite{Erro2010}, \cite{Tian2015}. By limiting the conversion to only warping the frequency axis of a high-dimensional source spectrum towards the target spectrum, spectral details are preserved. Nevertheless, the estimation of the exact warping functions is not straightforward \cite{Tian2014}. The newly-emerging exemplar-based approach \cite{Takashima2012}, \cite{NMF_Wu2014}, \cite{Wu2014} operates on high-dimensional features too. As a few basis spectral features are used and combined to form the output spectrum, the over-smoothing issue in GMM-based approach therefore no longer exists.

Recently, DNN-based techniques have been well presented in speech community \cite{YuDong2011}, \cite{ZHLing2015}. Voice conversion using DNN models generates non-linear mapping between source and target features, and there is little restriction in the feature dimensions to be modeled. A very early work on DNN-based spectral conversion focusing formant transformation was presented in \cite{narendranath1995transformation}. Some other pioneering works in spectral conversion include \cite{Desai2009} and \cite{Nakashika2013dbn}. Other works employed Restricted Boltzmann Machines (RBM) to estimate joint distribution \cite{ling2013joint}, \cite{wu2013conditional} or to estimate high level features \cite{Nakashika2014rtrbm}, \cite{chen2014multiframe}. A recent work on spectral conversion \cite{SequenceError_Xie2014} used DNN to perform transformation directly on high-dimensional spectral features, delivering accurate conversion and decent speech quality. Voice conversion using multiple speaker input has also been investigated in \cite{liu2015multispeaker}. Our proposed voice conversion on various prosodic and spectral features is built by leveraging on the above merits of DNNs.

Speaker characteristics are carried by various speech features. Besides spectrum, comprehensive voice conversion requires other features for transformation, for example, F0. Nevertheless, unlike spectrum, there are not many established works on F0 conversion. Some methods include transformation using vector quantization \cite{Helander2007}, partial least square regression \cite{Sanchez2014} and DNN mapping \cite{Pitch_Xie2014}. Among these, \cite{Sanchez2014} and \cite{Pitch_Xie2014} have focused on wavelet domain instead of on the frequency domain directly. The most popular method is the mean-variance global transformation, adjusting the average level and the range of source F0 values \cite{Stylianou1998}. This method retains the general shape of the source F0 trajectory and ignores the detailed difference between the two F0 contours. It is a frame-level operation, while human manipulates F0 in a segmental manner with a scope which may be up to phone, syllable, word, phrase or sentence level. F0 modeling and generation becomes challenging in voice conversion, as the amount of training data is usually sparse with tens to a few hundreds of utterances only.

Other prosody characteristics such as duration and intensity have been investigated in expressive speech synthesis and voice conversion. Representative works include duration modification based on HMM model \cite{Wu2006}, mean-variance transformation for phone or utterance duration \cite{Srikanth2012}, state duration modification using decision tree \cite{Yoshimura1999} and duration embedded GMM-based conversion \cite{Yutani2008}. Most of these methods use phone boundaries or phone identities from text labels. In \cite{Yutani2008}, consecutive spectral frames are used for duration modeling. Studies have also suggested some ways to perform intensity conversion based on F0 to improve naturalness \cite{Sorin2015}.

In this work, we explore the feasibility of building a comprehensive voice conversion framework. Specifically, our contributions include the following: (1) Various features with spectrum information and those with prosody information such as F0, intensity and phone duration are modified under a comprehensive DNN framework. Our spectral conversion operates directly on high-resolution full spectra and the frame-level mapping is estimated; (2) We propose a new pretraining scheme based on autoencoder, to further improve the training of the DNN model under the condition of limited training data in common voice conversion applications; (3) On one hand, unlike spectrum, prosodic features such as F0, intensity and duration depend very much on the semantic meaning and speaking style, which in turn depend on suprasegmental information \cite{Meyer1961}. On the other hand, most voice conversion systems work on limited number of training utterances, which makes it hard to model suprasegmental context. In this work, we move from the conventional frame-level modeling and propose a segment-level modeling scheme, aiming to balance between the context length and the amount of training data. Particularly, the input and output for F0 and intensity are based on every contiguous voiced segment, rather than a feature frame or short moving window; (4) Our F0 conversion is aimed at the overall F0 trajectory, instead of single F0 values. Direct modeling of F0 difference between adjacent frames is used to emphasize the F0 movement over time and consequently similar F0 trajectories can be compactly modeled. This facilitates the learning of DNN model parameters, which is crucial for the common scenario of limited voiced segments in training data of voice conversion. Finally, similar to our F0 modeling, duration ratio is modeled segmentally with input patterns constructed with sampled spectral frames.

The above proposed framework has been examined in details. The conversion performance delivered by various features was verified. In particular, the effectiveness of our autoencoder pretraining is assured and an outstanding performance in objective measurement was achieved. Our comprehensive conversion framework was shown to provide the highest rating on speaker similarity.

In the following sections, we will briefly review GMM-based and DNN-based voice conversion systems, which are two state-of-the-art and are also our baselines for evaluation.

\section{GMM-based Voice Conversion}
\label{sec:gmm}

Joint Density GMM (JD-GMM) \cite{Toda2007}, \cite{Kain1998} is one of the most successful methods in GMM-based voice conversion. In this method, GMM is estimated to model the joint distribution of source speech $\bm{X}$ and target speech $\bm{Y}$. During training, expectation-maximization (EM) method is used to maximize the joint probability as below:

\begin{gather}
P(\bm{X}, \bm{Y}) = P(\bm{Z}) = \sum_{k = 1}^K w_k \mathcal{N}(\bm{z}|\bm{\mu}_k^{(z)}, \bm{\Sigma}_k^{(z)}) \\
\bm{z} = \begin{bmatrix} \bm{x} \\ \bm{y} \end{bmatrix}, \quad \bm{\mu}_k^{(z)} = \begin{bmatrix} \bm{\mu}_k^{(x)} \\ \bm{\mu}_k^{(y)} \end{bmatrix}, \quad \bm{\Sigma}_k^{(z)} = \begin{bmatrix} \bm{\Sigma}_k^{(xx)} & \bm{\Sigma}_k^{(xy)} \\ \bm{\Sigma}_k^{(yz)} & \bm{\Sigma}_k^{(yy)} \end{bmatrix}
\label{eq:GMM_JD_Training}
\end{gather}
where $\bm{Z}$ is the joint paired feature vector sequence, $K$ is the number of Gaussian components, $w_k$ is the weight of the $k$-th component and $\mathcal{N}(\bm{z}|\bm{\mu}_k^{(z)}, \bm{\Sigma}_k^{(z)})$ is the Gaussian distribution with mean $\bm{\mu}_k^{(z)}$ and covariance matrix $\bm{\Sigma}_k^{(z)}$ of the $k$-th component.

During conversion, minimum mean square error is employed to estimate the target feature vector $\hat{\bm{y}}$ from each input source feature vector $\bm{x}$:

\begin{gather}
\hat{\bm{y}} = F(\bm{x}) = \sum_{k = 1}^K p_k(\bm{x}) \bigl [\bm{\mu}_k^{(y)} + \bm{\Sigma}_k^{(yx)}(\bm{\Sigma}_k^{(xx)})^{-1}(\bm{x} - \bm{\mu}_k^{(x)})\bigr ]
\label{eq:GMM_JQ_Conversion} \\
p_k(\bm{x}) = { w_k \mathcal{N}(\bm{x}|\bm{\mu}_k^{(x)}, \bm{\Sigma}_k^{(xx)}) \over \sum_{k = 1}^K w_k \mathcal{N}(\bm{x}|\bm{\mu}_k^{(x)}, \bm{\Sigma}_k^{(xx)})}
\end{gather}
where $p_k(\bm{x})$ is the posterior probability of the source vector $\bm{x}$ generated from the $k$-th Gaussian component.

\section{DNN-based Voice Conversion}
\label{sec:dnn}

\begin{figure}
\begin{center}
\includegraphics[width=300px]{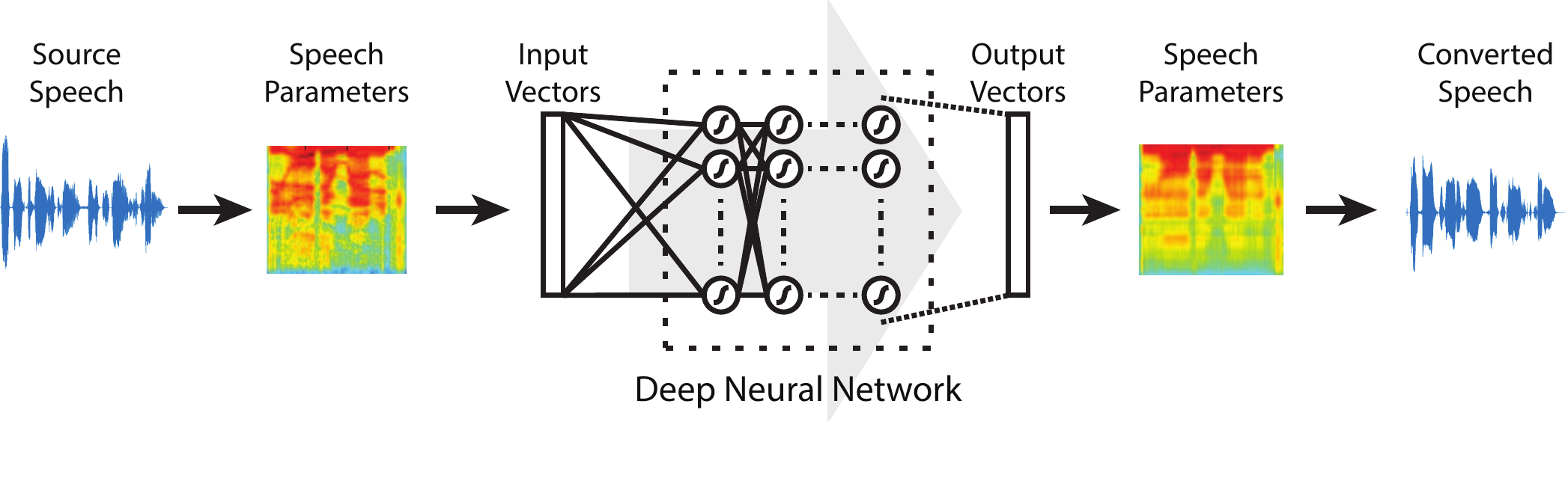}
\caption{DNN-based Voice Conversion}
\label{fig:dnn}
\end{center}
\end{figure}

In this approach, a DNN is used to model the transformation from source to target speech features as demonstrated in Fig.~\ref{fig:dnn}. Most of the state-of-the-art DNN-based voice conversion methods are on spectral transformation. In those methods, spectral parameters of the source speech are extracted and put into an input vector for conversion. This input vector is then passed to a DNN-based conversion network, where the network performance is governed by the network architecture, node weights and biases and the associated activation functions. By feeding forward the input vector through the network, an output vector is generated which represents the converted speech parameters. These output parameters are finally used to construct the converted waveform. Conversion on other features, such as F0, can also be done similarly.

However, the appropriate set of DNN model parameters that transforms source to target feature vectors is unknown at the beginning. A training phase is required. Before model training, random initialization on the model parameters is applicable or a pretraining step using layer-wise restricted Boltzmann machine (RBM) pretraining \cite{LHChen2014} or discriminative pretraining \cite{YuDong2015} can be performed. During model training, the error measurement of each hidden node is calculated using back-propagation algorithm \cite{Rumelhart1986}, and the DNN model parameters are updated accordingly to optimize a training criterion. A common training criterion in voice conversion is the sum of squared error between the target output vector and the model output. At the conversion phase, input vectors are propagated forward through the model with these estimated parameters to produce the corresponding output vectors.

One of the pioneering work in voice conversion using DNN is \cite{Desai2009} in which a low-dimensional spectral representation, mel-cepstral coefficients (MCEP), is used directly as input and output vectors for a feed forward DNN. Another work \cite{Nakashika2013dbn} employs Deep Belief Nets (DBNs) to extract  latent features from source and target cepstrum coefficients, and uses a neural network with one hidden layer to perform conversion between latent features. High-dimensional representation of spectrum has also been used in a more recent work \cite{SequenceError_Xie2014} for spectral mapping, together with dynamic features and parameter generation algorithm \cite{ParameterGeneration_Tokuda2000}. Moreover, DNN model has also been used to generate the F0 of target speaker with input spectral parameters from the source speaker \cite{Pitch_Xie2014}.

\section{Proposed Conversion Framework}
\label{sec:system}

\begin{figure}
\begin{center}
\includegraphics[width=0.75\textwidth]{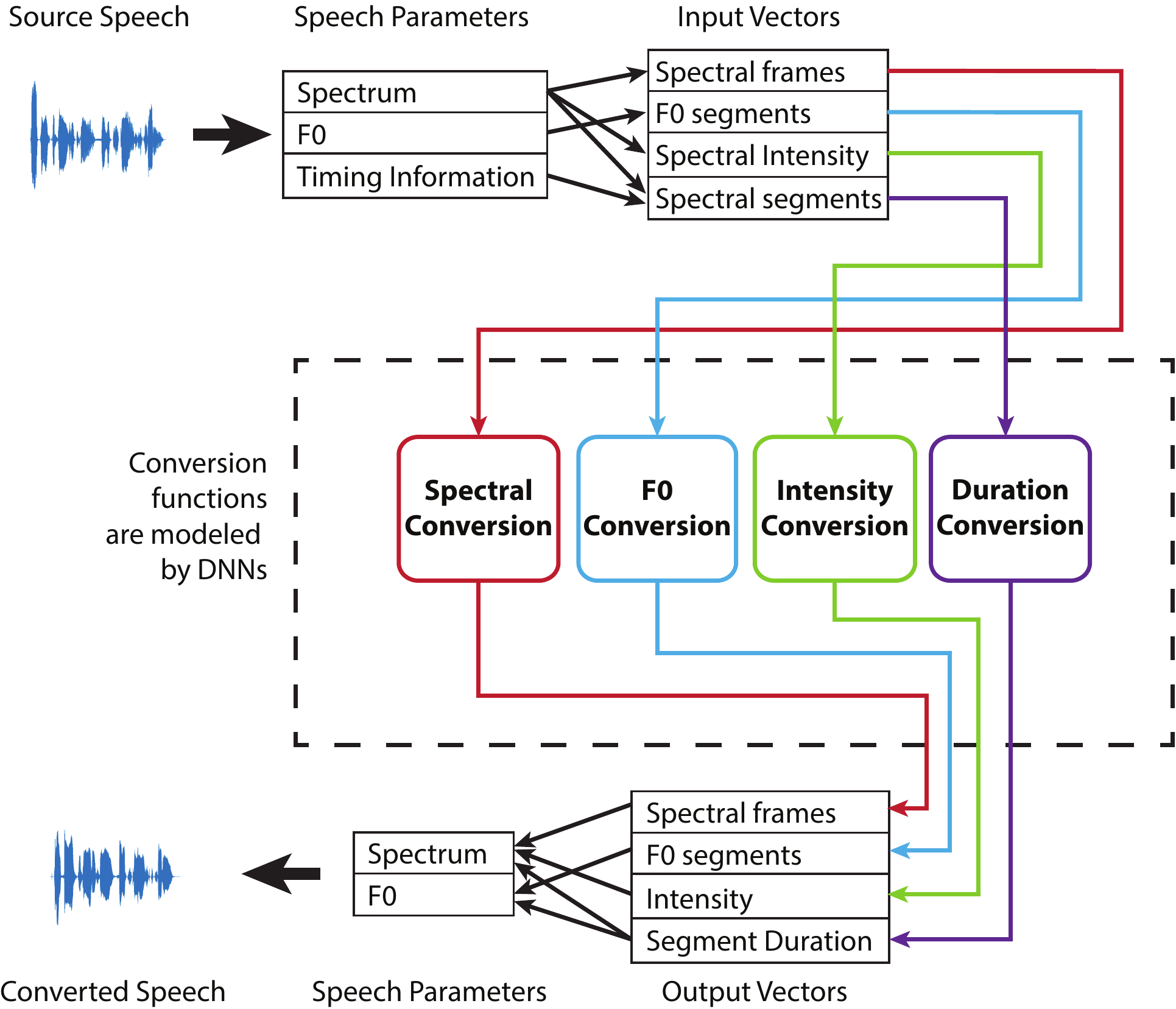}
\caption{Comprehensive Voice Conversion Framework}
\label{fig:framework}
\end{center}
\end{figure}

We now propose a comprehensive framework where spectrum, F0 contour, intensity trajectory and phone duration are all modeled using feed-forward DNNs. Due to distinct natures of various features (frame-level and segment-level), we used separate DNN model for each input feature.
The framework is demonstrated in Fig.~\ref{fig:framework}.

We extracted speech features from the input source speech and create appropriate input vectors for each conversion system. Each system is modeled by a DNN, of which parameters are estimated from the set of training data with the back-propagation algorithm and mean square error criteria. The converted outputs from each DNN model of individual features will be finally combined to reconstruct the output converted speech waveform. Detailed specifications of each model will be introduced below.

\subsection{Spectrum Conversion}
\label{sec:system.spectrum}
We use a feed-forward DNN to model the relationship between the source and target spectra. The DNN input is a source spectrum frame with dynamic features as context, while the DNN output is the corresponding target spectrum frame. In the current work, for simplicity, we do not employ parameter generation algorithm \cite{ParameterGeneration_Tokuda2000} with delta and delta-delta at output layer.

Our spectrum conversion process is partly based on \cite{SequenceError_Xie2014}. To improve the conversion performance further from the above DNN-based conversion framework, we have made two modifications described below.

\subsubsection{Two-stage Alignment}
\label{sec:2stage}

During data preparation, a two-stage alignment process is performed \cite{Yvonne2014}: Each training utterance are cut into phone segments using a phone recognizer in the first stage and then corresponding individual phone segments of source and target speaker are aligned using dynamic time warping (DTW) in the second stage. This aims to give a highly-accurate frame alignment, so that the DNN model training becomes more efficient.

\subsubsection{Autoencoder Pretraining}
\label{sec:autoencoder}

Pretraining plays an important role in DNNs \cite{ZHLing2015}. As the number of nodes in multiple layers ranges from hundreds to thousands, the resultant parameter space is enormous. Training a DNN with multiple hidden layers without pretraining is conventionally difficult. In the following, we propose a new pretraining technique in voice conversion by merely using the source spectral frames as both the input and output vectors and employing the L1 norm constraint on DNN weights. This is similar to training an autoencoder with sparsity constraint \cite{Le2011}. Autoencoder generally has much smaller number of nodes at the hidden layer in order to have dimension reduction and prevent the network from learning the identity function. However, as this autoencoder training acts as the pretraining for our spectral conversion DNN model, the hidden layers are not made to be narrow. Instead, we keep the same network architecture as of the DNN conversion network and use L1 norm constraint on DNN weights to employ a soft constraint on the sparsity of the activations of the hidden nodes.

This pretraining is motivated by the fact that both source and target spectral frames are aligned and in the same spoken content. The spectral shapes represented in the two features are roughly similar; the inter-difference may represent the speaker characteristics. Hence, by training an autoencoder of the source features, this provides a concise and effective initialization of the model parameters, which is believed to approximate the coarse spectral shape of the target feature. 

\begin{figure}
\begin{center}
\includegraphics{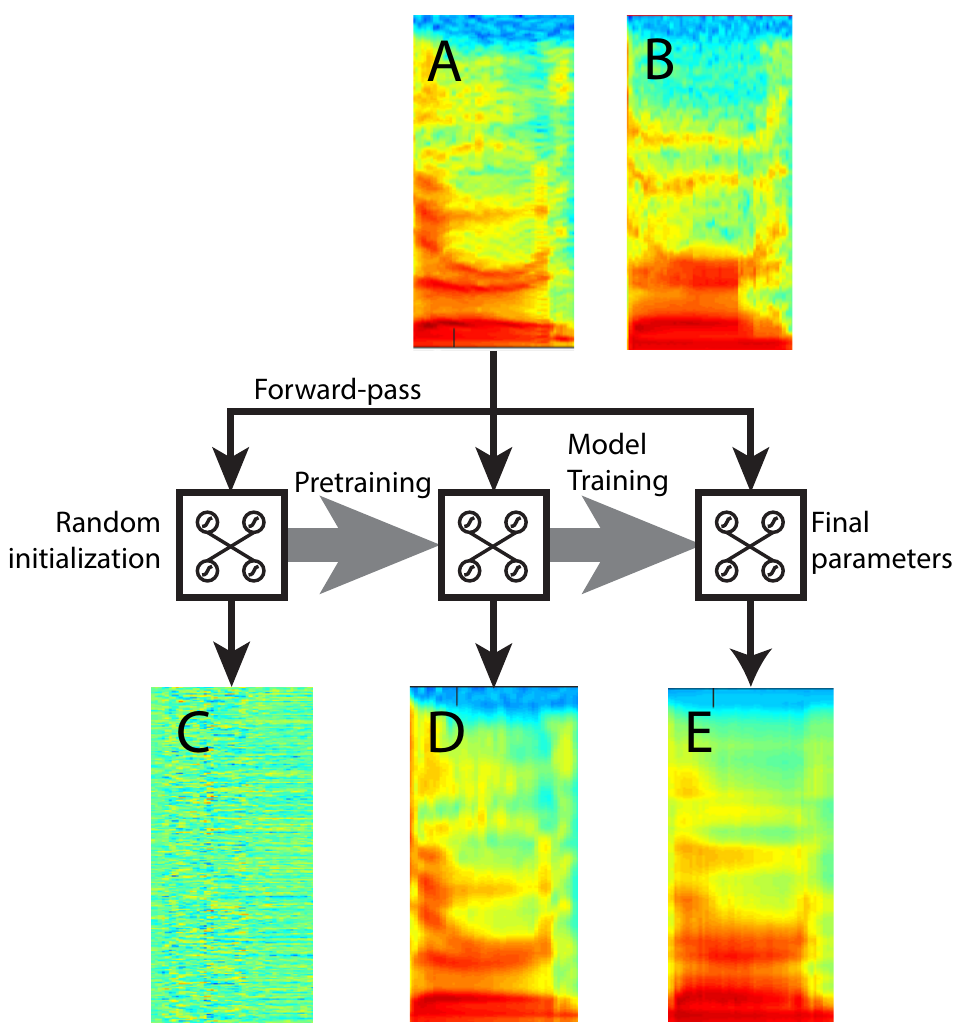}
\caption{Performing the forward-pass with the DNNs in different stages to demonstrate the effectiveness of autoencoder pretraining. A: spectrogram from source speech; B: spectrogram from target speech; C: output spectrogram from DNN with randomly initialized parameters; D: output spectrogram from DNN after autoencoder pretraining; E: output spectrogram from DNN after model training.}
\label{fig:autoencoder}
\end{center}
\end{figure}

At the beginning of this pretraining, all network weights are randomly initialized with uniform distribution and all biases are initialized at zero. The training criterion is the sum of squared error. After the pretraining error has converged, we remove the L1 norm constraint and use current network weights and biases as the initialization for the subsequent DNN model training on source and target features. 

Note that this autoencoder pretraining for voice conversion is different from the layer-wise autoencoder pretraining in \cite{Vincent2010}, where the autoencoder hidden representations given by the current layer are used as the input to the next layer. In our autoencoder pretraining for voice conversion, all parameters in the whole DNN are updated simultaneously. RBM pretraining is another popular pretraining method for DNN training. However, the optimization algorithm Contrastive Divergence \cite{hinton2002} that is used for RBM pretraining is unsupervised and not targeted on reconstructing spectral parameters. Moreover, comparing to discriminative layer-wise pretraining, autoencoder pretraining does not require early termination, which is usually required in discriminative layer-wise pretraining to prevent overfitting problem on each hidden layer.

In typical DNN training, there are three consecutive stages: initialization, pretraining and model training (fine-tuning). DNN-based voice conversion systems normally use only network after model training for conversion and evaluation. However, to demonstrate the effectiveness of the proposed pretraining, we would like to look at the output of the model in all the three stages, which are showed in Fig.~\ref{fig:autoencoder}. Fig.~\ref{fig:autoencoder}A and \ref{fig:autoencoder}B are the source and target spectrograms respectively. In Fig.~\ref{fig:autoencoder}C, the output spectrogram from a DNN with randomly initialized parameters loses all the spectral information, showing that the DNN exhibits a very inaccurate estimation of the spectral transformation function. In Fig.~\ref{fig:autoencoder}D, after the autoencoder pretraining step, the output spectrogram get much closer to the source spectrogram, which indicates a much more meaningful transformation represented by the DNN parameters. In Fig.~\ref{fig:autoencoder}E, after further training with target speech, the output spectrogram get closer to the target spectrogram in only a few number of epochs. In this example, we use 40 epochs for autoencoder pretraining and 20 epochs for model training.

\subsection{F0 Trajectory Conversion}
\label{sec:system.F0}

In our F0 conversion, we aim to model the transformation between the trajectories of source F0 and target F0. Since F0 is a prosodic feature that generally depends on long context information \cite{Meyer1961}, we performed a segment-level conversion instead of frame-level scheme as in the above spectrum conversion and other typical DNN-based voice conversion \cite{Pitch_Xie2014}. 

The segments are defined by the voiced-unvoiced boundaries, where each segment is a continuous F0 measurement in voiced frames. The length of those segments are first normalized into a predefined number $L$ before the segments are passed to the DNN model. The length normalization is done by interpolating the extracted F0 segment into a fixed length (This normalized length can be chosen empirically from the distribution of segment length observed in training data).  Given the suprasegmental nature of F0 and the training amount of voice conversion, in particular the amount of voiced segments, is usually small, this length normalization is important that F0 trajectory is concisely captured and compact modeling is facilitated.

To model the F0 contour, rather than the F0 values itself, our feature vector for modeling $t'_{i, j}$ is formed by,
\begin{equation}
t'_{i, j} = \begin{cases}
    0, & \text{if $i = 1$}\\
    t_{i, j} - t_{i-1, j}, & \text{if $i > 1$}.
  \end{cases}
\label{eq:F0_difference_modeling_target_feature_vector}
\end{equation}
where $t{'}_{i, j}$ is the $i$-th element of the $j$-th modeled feature vector. $t_{i, j}$ is the $i$-th element in the $j$-th length-normalized F0 segment of the target speaker. This is motivated by our previous findings on F0 modeling that relative pitch feature is effective for singing voice synthesis \cite{Yvonne2012} and small amount of training data: Multiple segments could have different absolute F0 levels while their F0 trajectories are similar. Modeling the F0 contour allows those segments to share model parameters, which would alleviate the limitation of small amount of training data.

The resultant model output is an F0 trajectory, such that a reference mean level of the target F0 segment is necessary for the final F0 reconstruction. In this work, this target segmental mean level is calculated using the converted segment value from a frame-based F0 conversion system \cite{Pitch_Xie2014}. Other ways of calculating the segmental mean level are also possible.
The reconstruction process from output segments is done in following three steps:
\begin{enumerate}
	\item An F0 trajectory is reconstructed from output feature vector, assuming that the first value is 0.
	\begin{equation}
	\hat{t}_{i, j}= \begin{cases}
	0, & \text{if $i = 1$}\\
	\hat{t}_{i - 1, j} + \hat{t}'_{i, j}, & \text{if $i > 1$}.
	\end{cases}
	\end{equation}
	where $\hat{t}'_{i, j}$ is the $i$-th element in the $j$-th output segment from the DNN model.
	
	\item Mean adjustment is applied to each reconstructed trajectory
	\begin{equation}
	\hat{t}_{i, j}^{*} = \hat{t}_{i, j} - {\sum_{i=1}^{L} \hat{t}_{i, j} \over L} + \hat{\mu}_j
	\end{equation}
	where $\hat{\mu}_{j}$ is the predicted segmental mean level of the $j$-th segment and $\hat{t}^{*}_{i, j}$  is $i$-th element of the $j$-th the mean-adjusted F0 segment.
	
	\item The resultant segment is interpolated again into original length.
\end{enumerate}

\subsection{Intensity Trajectory and Phone Duration Conversion}
\label{sec:system.intensity_duration}

For intensity, we perform a similar segmentation process as in F0 trajectory conversion, whereas intensity trajectories in voiced segments are extracted and then normalized in length. However, we choose to use intensity values directly in input and output feature vectors for DNN training in this work. Based on our informal experiments, similar performance is observed when using the intensity ratio or difference between consecutive frames as the input features. At conversion, converted spectrum will be scaled using the predicted intensity trajectory.

\begin{figure}
\begin{center}
\includegraphics[width=0.75\textwidth]{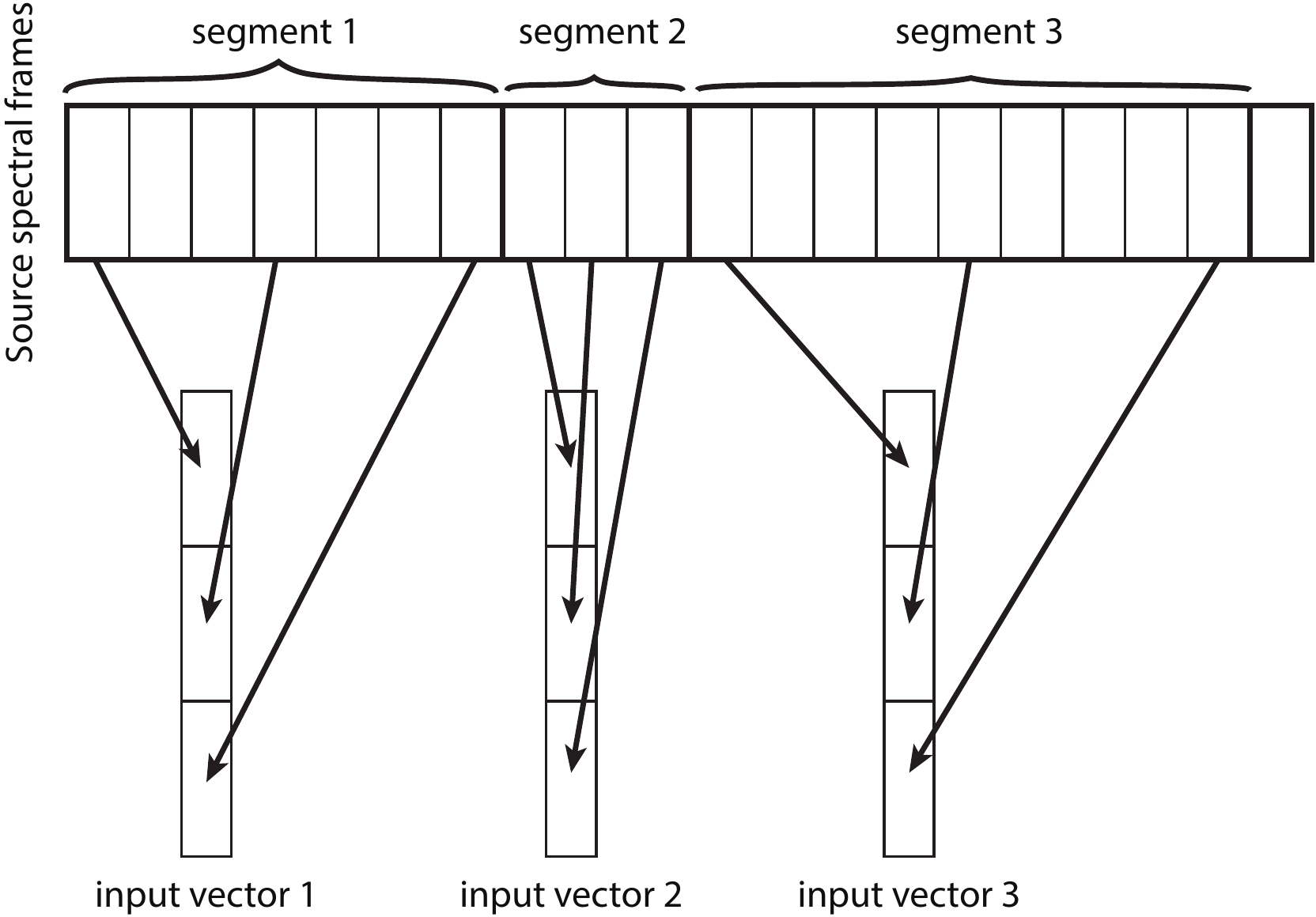}
\caption{Re-sample spectral segment to create input vector for duration modeling}
\label{fig:durationresample}
\end{center}
\end{figure}

For duration conversion, we aim to model the duration ratio between the source and target segment. By segment here, we refer to a contiguous spectral block associated with a phone, a syllable, etc. The input feature is constructed by sampling spectral frames from this source segment. This is demonstrated in Fig.~\ref{fig:durationresample}. In this example, each segment is constituted by three sampled frames. The number of sampled frames can be chosen empirically from the distribution of segment duration in the training data.

Without duration modification, all converted features has the same time alignment as the corresponding utterances from source speaker. Hence, we can employ the original segment timing of source speaker and the predicted phone duration ratio from the DNN model to modify the segment length by performing linear interpolation on all converted features. In this work, we explore the performance of duration modeling with phone-based segment units.

\section{Experiments}
\subsection{Experimental Setup}
\label{sec:experimental_setup}

To evaluate the performance of the proposed works, we carried out several experiments on the CMU ARCTIC database \cite{Kominek03cmuarctic}. In our experiments, we used four speaker pairs:
\begin{itemize}
	\item BDL (male) to SLT (female)
	\item CLB (female) to RMS (male)
	\item CLB (female) to SLT(female)
	\item RMS (male) to BDL (male)
\end{itemize}

200 utterances in the dataset were used as training set and 60 other utterances were used as testing set. We used TANDEM-STRAIGHT \cite{Kawahara2008} to calculate 512th-order log spectral envelope, voiced-unvoiced flag (VUV) and F0 with 5ms frame shift and 1024-point FFT. These speech features were then used to create different input and output vectors for our conversion models.

The detailed descriptions of each input, output features and network architectures are described below:
\begin{itemize}
	\item \textbf{Spectrum conversion}: We used 512th-order log spectral envelope and its delta and delta-delta coefficients and one dimension of VUV from source speaker as input feature vector and 512th-order log spectral envelope from target speaker as output feature vector. The DNN had three hidden layers; each hidden layer had 3000 nodes.
	\item \textbf{F0 trajectory conversion}: We extracted F0 input and output segments as described in Section~\ref{sec:system.F0} with the normalized segment length $L$ of 55. The DNN had two hidden layers; each hidden layer had 500 nodes.
	\item \textbf{Intensity trajectory conversion}: Input and output vectors extracted as described in Section~\ref{sec:system.intensity_duration} with the normalized segment length of 55. The DNN architecture is the same as in F0 trajectory conversion stated above.
	\item \textbf{Phone duration conversion}: For each phone in source utterances, an input vector was created by concatenating five spectrum frames from the phone, and the output vector was the ratio between the current source phone duration and the corresponding target phone duration. The phone boundaries can be determined by any phone recognizer. In our work, we use the boundaries from CMU Sphinx recognizer \cite{lamere2003design}.
\end{itemize}

All the DNNs in this work employed hyperbolic tangent activation in the hidden nodes and there was no non-linear activation function in any output nodes. In all the system below, aperiodicity feature was not converted. We use the same aperiodicity feature from the source speaker to reconstruct the converted speech.

\subsection{Spectrum Conversion}

We evaluate the performance of DNN-based spectral conversion using the following systems:
\begin{itemize}
	\item \textbf{JD-GMM}: The spectral representation was 25-th order MCEP, and the training and conversion was done as described in Section~\ref{sec:gmm}. We used 64 Gaussian components to build the system, which is empirically found in our preliminary experiments.
	\item \textbf{DNN-MCEP}: This system also used 25-th order MCEP as source and target feature. For DNN architecture and training parameters, we used the same settings as in \cite{Desai2009}: two hidden layers with 50 nodes each layer, the learning rate was 0.01 and the momentum was 0.3.
	\item \textbf{DNN-SP256-DLP}: We built a system based on \cite{SequenceError_Xie2014}, in which the input vectors include log spectral frames with delta and delta-delta coefficients, log F0 with delta and delta-delta of the current frames and two context frames and one dimension of VUV. The DNN had two hidden layers with 1600 nodes each layer. We applied discriminative layer-wise pretraining (DLP) on this system.
	\item \textbf{DNN-SP512} (proposed): We used a DNN with the architecture described in Section~\ref{sec:experimental_setup}. We did not perform any pretraining on this system.
	\item \textbf{DNN-SP512-DLP} (proposed): We used a DNN with the architecture described in Section~\ref{sec:experimental_setup} and applied discriminative layer-wise pretraining (DLP) on this system.
	\item \textbf{DNN-SP512-Autoencoder} (proposed): We used a DNN with the architecture described Section~\ref{sec:experimental_setup} and performed autoencoder pretraining as described in Section~\ref{sec:system.spectrum}. The training data for the pretraining phase were 200 utterances from source speaker. This same set was also used in the model training stage.
\end{itemize}

\subsubsection{Objective evaluation}
\label{sec:sp.obj_eval}

We conducted objective evaluation to assess the proposed spectral conversion with autoencoder pretraining. Log spectral distortion (LSD) \cite{ye2004high} was employed. The distortion of $k$-th order of log spectral pair is calculated as:

\begin{equation}
d(\bm{x}_k, \bm{y}_k) = \sum_{i=1}^M(log(x_{k,i}) - log(y_{k,i}))^2
\end{equation}
where $M$ is the total number of frequency bin. A distortion ratio between converted-to-target distortion and the source-to-target distortion is defined as:

\begin{equation}
\text{LSD} = {\sum_{k=1}^K d(\hat{\bm{x}}_k, \bm{y}_k) \over \sum_{k=1}^K d(\bm{x}_k, \bm{y}_k)} * 100\%
\end{equation}
where $\bm{x}_k$, $\bm{y}_k$ and $\hat{\bm{x}}_k$ denote the $k$-th frequency bin of the source, target and converted spectrum, respectively, and $K$ denotes the number of frequency bins. A lower LSD indicates smaller distortion.

\begin{table}
\begin{center}
\caption{Comparison of LSD ratio of different spectral conversion methods}
\label{tab:sp.lsd}
\begin{tabular}{ccc} \hline \noalign{\smallskip}
Methods & LSD (\%) \\ \hline \hline \noalign{\smallskip}
JD-GMM & 87.93 \\
DNN-MCEP & 80.70 \\
DNN-SP256-DLP & 52.51 \\ \hline \noalign{\smallskip}
DNN-SP512 & 54.73 \\ 
DNN-SP512-DLP & 52.59 \\ 
\textbf{DNN-SP512-Autoencoder} & 51.31 \\ \hline
\end{tabular}
\end{center}
\end{table}

Table~\ref{tab:sp.lsd} presents the LSD for the baseline methods and our proposed methods. The average LSD over all evaluation pairs was reported. Comparing JD-GMM and DNN-MCEP with other methods, we first observed that the systems working on high dimensional representations of spectrum (i.e. DNN-SP256-DLP, DNN-SP512, DNN-SP512-Autoencoder) yield much lower distortion than those working on low dimensional representations (at least 25\% lower). 

Comparing among DNN-SP256-DLP, DNN-SP512 and DNN-SP512-DLP, DNN-SP512 and DNN-SP512-DLP achieve a slightly higher distortion than DNN-SP256-DLP, that is 54.73\% and 52.59\% to 52.51\%, which might indicate the difficulty in achieving a good minimum in training DNN with high dimensional features. However, when employing our proposed autoencoder pretraining, DNN-SP512-Autoencoder converges to a better solution with a lower LSD, and has a slightly lower distortion than DNN-SP256-DLP, that is 51.31\% to 52.51\%. This indicates that the proposed autoencoder pretraining enables a better initialization for spectral conversion.

Note that discriminative layer-wise pretraining was applied in DNN-SP256-DLP; while in our proposed DNN-SP512-Autoencoder, autoencoder pretraining was implemented. Comparing the LSD performance of these two pre-trained methods, the combination of autoencoder pretraining and DNN-SP512 (DNN-SP512-Autoencoder) yields a lower LSD with a doubled resolution in frequency.

\subsubsection{Subjective evaluation}
\label{sec:sp.sub_eval}

We conducted listening tests to access speech quality and speaker similarity of the converted results. 10 subjects participated in all listening tests. In this subjective evaluation, we compared DNN-SP512-Autoencoder with DNN-SP256-DLP, DNN-MCEP and JD-GMM. We did not include DNN-SP512 nor DNN-SP512-DLP due to the highly similar structure with DNN-SP512-Autoencoder.

We first performed AB preference tests to access speech quality. 20 pairs were randomly selected from 80 paired samples. In each pair, A and B were the samples from the proposed method and a baseline method, respectively, presented to listeners in a random order. Each listener was asked to listen to both samples and decide which sample was better in term of quality.

We then conducted XAB tests to access the speaker similarity. Similar to the AB test, 20 pairs were randomly selected from the 80 paired samples. In each pair, X was the reference target sample, and A and B were the converted samples of comparison methods listed in the first column of Table~\ref{tab:sp.subjective}, presented to listeners in a random order. The listeners were asked to listen to the sample X first, then A and B, and then decide which sample was closer to the reference target sample. For each pair of samples, X, A and B have the same language content.

\begin{table}
\begin{center}
\caption{Results of average quality and similarity preference tests with 95\% confidence intervals of different spectral conversion methods}
\label{tab:sp.subjective}
\begin{tabular}{ccc} \hline \noalign{\smallskip}
\multirow{2}{*}{Methods} & \multicolumn{2}{c}{Preference score (\%)} \\ \cline{2-3} \noalign{\smallskip}
& Quality test & Similarity test \\ \hline \hline \noalign{\smallskip}
JD-GMM & 35.8 ($\pm$ 6.7) & 36.8 ($\pm$ 8.3) \\
\textbf{DNN-SP512-Autoencoder} & 64.2 ($\pm$ 6.7) & 63.2 ($\pm$ 8.3) \\ \hline \noalign{\smallskip}
DNN-MCEP & 42.5 ($\pm$ 3.8) & 42.0 ($\pm$ 6.6) \\
\textbf{DNN-SP512-Autoencoder} & 57.5 ($\pm$ 3.8) & 58.0 ($\pm$ 6.6) \\ \hline \noalign{\smallskip}
DNN-SP256-DLP & 46.6 ($\pm$ 3.8) & 44.0 ($\pm$ 6.4) \\
\textbf{DNN-SP512-Autoencoder} & 53.3 ($\pm$ 3.8) & 56.0 ($\pm$ 6.4) \\ \hline
\end{tabular}
\end{center}
\end{table}

The subjective evaluation results are presented in Table~\ref{tab:sp.subjective}. Comparing to JD-GMM, our proposed method DNN-SP512-Autoencoder achieves much higher preference score in both quality and similarity test. Comparing to DNN-MCEP, DNN-SP512-Autoencoder also yields better results in both quality and similarity evaluations. Note that the spectral resolution in DNN-SP512-Autoencoder is much higher than JD-GMM and DNN-MCEP (In JD-GMM and DNN-MCEP, spectral envelope is represented by 25th-order MCEP). Spectral details are clearer in DNN-SP512-Autoencoder.

In the comparison between the state-of-the-art DNN-SP256-DLP and our proposed method DNN-SP512-Autoencoder, our proposed method achieves a slightly better subjective results in both quality and similarity. The results further confirms that DNN is capable of modeling high-resolution spectra and autoencoder pretraining can effectively improve the performance of DNN training for voice conversion.

\subsection{Spectrum and F0 Conversion}

We compared our proposed segment-based F0 conversion system with two baseline systems:
\begin{itemize}
	\item \textbf{Mean-Var}: F0 is converted using the conventional global mean-variance transformation from source F0 \cite{Stylianou1998} as follows,
\begin{equation}
f_{o} = \mu_{t} + {\sigma_{t} \over \sigma_{s}} * (f_{s} - \mu_{s}).
\label{eq:F0_mean_var_global_transformation}
\end{equation}
where $f_s$ and $f_o$ are the source F0 and the output converted F0, respectively. $\mu_{s}$, $\sigma_{s}$, $\mu_{t}$, $\sigma_{t}$ are the global mean and standard deviation statistics of the F0s from source and target speaker, respectively. 
	\item \textbf{DNN-Frame}: F0 is modeled using a frame-based approach with DNN \cite{Pitch_Xie2014}. In this system, the input vector contains 256th-order log spectral envelope (static, delta and delta-delta), F0 (static values within a 7-element window, delta and delta-delta) and VUV. The DNN has two hidden layers with 1600 nodes each.
	\item \textbf{DNN-Segment} (proposed): The F0 segment is modeled as described in Section~\ref{sec:system.F0}. The DNN architecture is described in Section~\ref{sec:experimental_setup}.
\end{itemize}

In all F0 conversion systems, the result from DNN-SP512-Autoencoder was used as spectrum feature for speech reconstruction.

\subsubsection{Objective evaluation}

We employed root mean square error (RMSE) as the objective measurement to compare our proposed methods with the baseline systems. The average RMSE over all evaluation pairs was reported. A lower RMSE indicates smaller distortion.

\begin{table}
\begin{center}
\caption{Comparison of RMSE of different F0 conversion methods}
\label{tab:f0.rmse}
\begin{tabular}{ccc} \hline \noalign{\smallskip}
Methods & RMSE (Hz) \\ \hline \hline \noalign{\smallskip}
Mean-Var & 22.06 \\
DNN-Frame & 17.90 \\ \hline \noalign{\smallskip}
\textbf{DNN-Segment} & 17.80 \\ \hline
\end{tabular}
\end{center}
\end{table}

Table~\ref{tab:f0.rmse} presents the RMSE for the baseline methods and our proposed method. Comparing to Mean-Var, DNN-Segment achieves lower RMSE, that is 17.80 to 22.06, and comparing to DNN-Frame, DNN-Segment achieves a slightly lower RMSE, that is 17.80 to 17.90, which confirms the effectiveness of the segment-based trajectory modeling method.

\subsubsection{Subjective evaluation}

We performed further evaluation with subjective listening test. We conducted an XAB similarity preference test in this evaluation. The setup is similarly to the XAB test in Section~\ref{sec:sp.sub_eval}. We instructed the listeners to choose the sample whose prosody sound more similar to the reference target sample.

\begin{table}
\begin{center}
\caption{Results of average similarity preference tests with 95\% confidence intervals of different F0 conversion methods}
\label{tab:f0.subjective}
\begin{tabular}{cc} \hline \noalign{\smallskip}
\multirow{2}{*}{Methods} & Preference score (\%) \\ \cline{2-2} \noalign{\smallskip}
& Similarity test \\ \hline \hline \noalign{\smallskip}
Mean-Var & 38.8 ($\pm$ 2.1) \\
\textbf{DNN-Segment} & 61.2 ($\pm$ 2.1) \\ \hline \noalign{\smallskip}
DNN-Frame & 48.8 ($\pm$ 4.1) \\
\textbf{DNN-Segment} & 51.2 ($\pm$ 4.1) \\ \hline
\end{tabular}
\end{center}
\end{table}

The subjective evaluation results are presented in Table~\ref{tab:f0.subjective}. The similarity tests further confirm that our proposed method DNN-Segment provides better F0 modeling than the conventional Mean-Var method by a significantly better preference score, that is 61.2 to 38.8. Our proposed segment-based method DNN-Segment also has comparable preference score with state-of-the-art F0 modeling using DNNs.

\begin{figure}
\begin{center}
\includegraphics[width=180px]{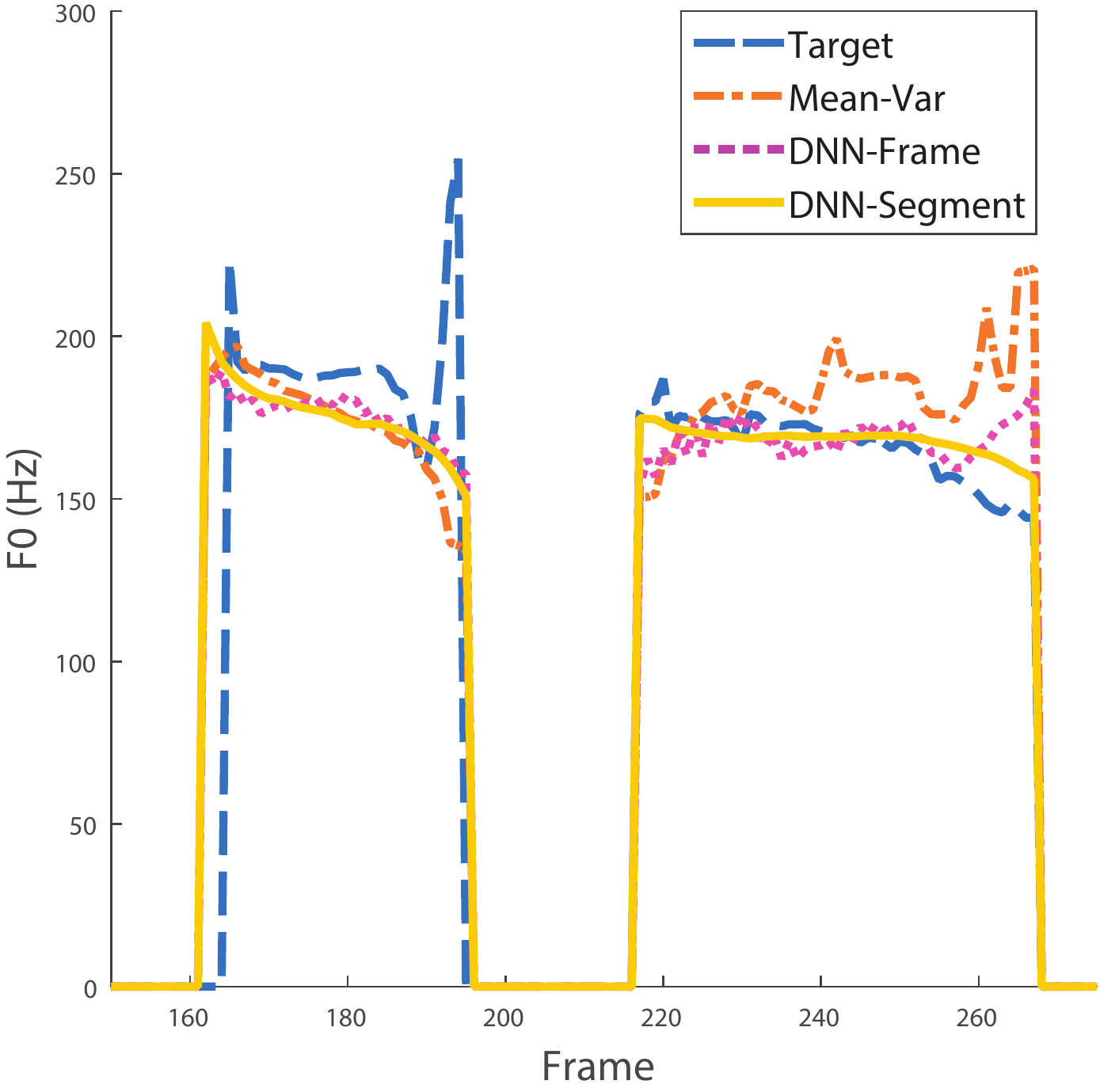}
\caption{Outputs of F0 conversion from three methods: Mean-Var, DNN-Frame and DNN-Segment}
\label{fig:F0sample}
\end{center}
\end{figure}

A demonstration of the results from F0 conversion method is showed in Fig.~\ref{fig:F0sample}.

\subsection{The Comprehensive Framework: Spectrum, F0 and Intensity and Duration Conversion}

Finally, we combined our spectrum conversion and F0 conversion system with the proposed intensity and duration conversion described in Section~\ref{sec:system.intensity_duration}. We compared the speaker similarity between the system with and without intensity and duration modification using listening tests. In this evaluation, we performed the evaluation on speaker pair CLB to SLT, where the intensity contour and time alignment between source and target are noticeably different.

\begin{table}
\begin{center}
\caption{Results of average similarity preference tests with 95\% confidence intervals of the proposed intensity and duration conversion method}
\label{tab:intensity_duration.subjective}
\begin{tabular}{cc} \hline \noalign{\smallskip}
\multirow{2}{*}{Methods} & Preference score (\%) \\ \cline{2-2} \noalign{\smallskip}
& Similarity test \\ \hline \hline \noalign{\smallskip}
Without intensity and duration & 31.2 ($\pm$ 4.1) \\
\textbf{With intensity and duration} & 68.8 ($\pm$ 4.1) \\ \hline
\end{tabular}
\end{center}
\end{table}

The subjective evaluation results are presented in Table~\ref{tab:intensity_duration.subjective}. The preference scores of systems with and without intensity and duration conversion are 68.8 and 31.2, respectively. This further suggests the effectiveness of our proposed system in converting prosodic features and the importance of having a comprehensive framework in voice conversion. 

Some samples of this work are presented in the web link: \url{http://listeningtests.net/voiceconversion/hynq2015comprehensive}.

\section{Conclusions}

This paper showcases a comprehensive voice conversion framework, in which high-resolution spectra, F0, intensity and segment duration are all converted using DNNs. The objective and subjective evaluations shows the capability of the modeling of high-dimensional full spectra and prosodic segments. The proposed autoencoder pretraining for voice conversion is shown to effectively initialize the model parameters and leads to accurate spectral estimation. The resultant spectral conversion model with our proposed pretraining achieves comparable ratings in terms of similarity and quality as another state-of-the-art pretrained voice conversion system \cite{SequenceError_Xie2014}. 

In conventional works of voice conversion, prosodic features are not commonly modeled. Knowing that human manipulates prosody in various levels, rather than in a local frame basis, we have introduced segment-level modeling and conversion for F0, intensity and duration, without increasing the amount of training data of typical voice conversion scenarios. Our experimental results have shown that the resultant converted speech is much similar to the target speaker. Throughout this work, a feasible conversion framework is built where multiple features are transformed acoustically and future investigations on ways of modeling and conversion of timbre and prosodic features remain thought-provoking. To better understand the performance of intensity and duration conversion, we plan to evaluate the perceptual effect of intensity and duration modification on voice conversion through more intensive psychoacoustic experiments.

\begin{acknowledgements}
This research is supported by the National Research Foundation, Prime Minister’s Office, Singapore under its IDM Futures Funding Initiative and administered by the Interactive and Digital Media Programme Office.
\end{acknowledgements}

\bibliographystyle{spmpsci}      
\bibliography{hybib}   

\end{document}